\documentclass[10pt,preprint]{aastex}
\usepackage{psboxit}

\newcommand{\kms}{\,km~s$^{-1}$}

\PScommands

\newcommand{\mdmlimit}{0.5}
\newcommand{\vmaxmean}{56}
\newcommand{\mrmean}{-14.7}

\def\simless{\mathbin{\lower 3pt\hbox
	{$\,\rlap{\raise 5pt\hbox{$\char'074$}}\mathchar"7218\,$}}} % < or of order
\def\simgreat{\mathbin{\lower 3pt\hbox
	{$\,\rlap{\raise 5pt\hbox{$\char'076$}}\mathchar"7218\,$}}} % > or of order

\newcounter{thefigs}
\newcommand{\fignum}{\arabic{thefigs}}

\newcounter{thetabs}

\newcounter{address}

\shortauthors{Blanton {\it et al.} (2007)}
\shorttitle{Galaxy mass function}

\begin{document}

\title{ Testing cold dark matter 
with \\
the low mass Tully-Fisher relation}

\author{
Michael R. Blanton\altaffilmark{\ref{NYU}},
Marla Geha\altaffilmark{\ref{NRC}}, and 
Andrew A.~West\altaffilmark{\ref{Berkeley}}}

%\altaffiltext{1}{Based on observations obtained with the
%Sloan Digital Sky Survey\label{SDSS}} 
\setcounter{address}{1}
\altaffiltext{\theaddress}{
	\stepcounter{address}
	Center for Cosmology and Particle Physics, Department of Physics, New
	York University, 4 Washington Place, New
	York, NY 10003
	\label{NYU}}
\altaffiltext{\theaddress}{
	\stepcounter{address}
	National Research Council of Canada, Herzberg Institute of
	Astrophysics, 5071 West Saanich Road, Victoria, BC V9E 2E7, Canada
	\label{NRC}}
\altaffiltext{\theaddress}{
	\stepcounter{address}
	Astronomy Department, University of
California, 601 Campbell Hall, Berkeley, CA 94720
	\label{Berkeley}}

\begin{abstract}
In most cosmological theories, the galaxy mass function at
small masses is related to the matter power spectrum on
small scales. The circular velocity function (a quantity
closely related to the mass function) is well-studied for
dwarf satellites in the Local Group. However, theoretical
predictions and observational measurements are difficult for
satellite galaxies, because of ram pressure and tidal
stripping. By contrast, isolated dwarf galaxies are less
affected by these processes, and almost always have enough
21cm emission to trace their dynamics robustly.  Here, we
use isolated low mass dwarf galaxies from the Sloan Digital
Sky Survey (SDSS), with measured 21cm widths, to test cold
dark matter cosmology.  We find consistency between the
predicted and observed number density of isolated galaxies
down to $V_{\mathrm{max}} \sim 50$ km s$^{-1}$.  Our
technique yields a direct test of small-scale cosmology
independent of the Lyman-$\alpha$ forest power spectrum, but
our sample is currently statistically less powerful: warm
dark matter particles heavier than $\mdmlimit$ keV cannot be
ruled out.  Our major systematic uncertainty is the surface
brightness limit of the SDSS. Blind HI surveys, such as the
ALFALFA survey on Arecibo, are expected to uncover a much
larger number of isolated low mass galaxies, will increase
the power of our constraints at small scales, and will
propel the study of isolated galaxies to low masses
previously attainable only in the Local Group. We use our
sample to explore dwarf galaxy formation as well, finding
that the Tully-Fisher relation for dwarf galaxies is a
strong function of environment, and that the baryon (stellar
plus neutral gas mass) fraction is only a weak function of
galaxy mass. Together with the strong dependence of gas
fraction on environment, these results indicate that for
dwarf galaxies, gas loss and the end of star-formation are
dominated by external, not internal, processes.
\end{abstract}


\section{ Introduction: the mass function and cosmology}
\label{intro}

A critical test for any theory of cosmology and structure formation is
whether it correctly predicts the galaxy mass function.  The current 
Cold Dark Matter model with a cosmological constant ($\Lambda$CDM;
\citealt{spergel06a}) makes robust predictions for the number of dark
matter halos as a function of mass, finding roughly that $dN/dM
\propto M^{-1.8}$ (\citealt{sheth01a,jenkins01a, reed03a,yahagi04a}).
A similar mass function describes the distribution of surviving
subclumps within the dark matter halos (``subhalos'') that most
investigators associate with the sites of galaxy formation
(\citealt{colin99a, gao04a, kravtsov04b, reed05a, zentner05a,
conroy05b}).  Of course, extending this model to predict the number
density of galaxies in the Universe requires including physical
effects such as gas cooling, star-formation, supernova feedback, and
possibly the formation of supermassive black holes and their feedback,
that are too complex to follow exactly in numerical
predictions. Nevertheless, in the last three decades numerous
approximate approaches to the problem have lent us some understanding
of what the correct predictions might be and what physical processes a
successful model may involve (e.g.,
\citealt{ostriker77a, white78a, white91a, blanton99a, somerville01a,
benson03a, robertson05a, croton06a}).

Recently, various investigators have tested the halo mass function at
high masses and found consistency with predictions based on Wilkinson
Microwave Anisotropy Probe three-year (WMAP3) cosmological parameters
(\citealt{bahcall03a, rines06a, eke06a}).  
%In fact, these sorts of
%studies preferred a low value of $\sigma_8$ ($\sim 0.8$ if $\Omega_m
%\sim 0.3$) before the release of the WMAP three-year results
%(\citealt{bahcall03a}).  
In addition, for individual, high luminosity galaxies one can perform
a similar test.  If galaxy luminosity is related monotonically (with
some scatter) to halo mass, then $\Lambda$CDM predicts the
weak-lensing signal as a function of galaxy
abundance. \citet{tasitsiomi04a} have verified this prediction for
galaxies with $L_r > L_{r, \ast}$, and \citet{seljak05a} have
performed a similar test extending down to about $0.1L_{r, \ast}$.

Below $0.1L_{r,\ast}$, the luminosity function is closer to a power
law, with a slope that varies somewhat over luminosity but is never
steeper than about $N \propto L^{-1.5}$ at most, significantly
shallower than the prediction for the number as a function of halo
mass (\citealt{trentham05a, blanton04b}).  If galaxies and subhalos
are associated one-to-one, then the mass-to-light ratios of galaxies
must increase substantially with decreasing luminosity. However, data
on galaxies at the lowest luminosities has in the past been rather
scarce. At circular velocities below $L_r \sim 0.1 L_{r_\ast}$ nobody
has demonstrated consistency between the $\Lambda$CDM prediction and
the galaxy mass function. Some valiant attempts exist, but rely on
extrapolating the \citet{tully77a} and \citet{faber76a} relations into
the low luminosity regime (\citealt{desai04a, goldberg05a}).

The most extreme example of this issue is the ``substructure problem''
in the Local Group (\citealt{klypin99b, kravtsov04a}). $\Lambda$CDM
predicts hundreds of low mass satellites of the Milky Way, but only
approximately twenty are known (though the number is growing month by
month; \citealt{ibata95a, mateo98a, willman05b, willman05a,
belokurov06a, zucker06a, zucker06b}). \citet{simon07a} show that
although the newly discovered galaxies have eased this discrepancy, a
factor of 2--4 difference remains even when reasonable baryonic
physics has been taken into account.  This test of cosmology probes
the lowest mass galaxies possible, and so is extremely sensitive to
differences in the mass function slope.  However, both predictions and
observations are quite difficult in the vicinity of a luminous
galaxy. From the point of view of theory, a number of processes, such
as tidal stripping, ram pressure stripping, dynamical friction, and
merging onto the large galaxy, can occur near luminous galaxies
(\citealt{kravtsov04a, bullock05a, zentner05a, mayer06a}).  When a
dwarf galaxy enters the environment of a large galaxy, tidal stripping
can reduce its mass. In addition, the larger dwarf galaxies are
preferentially dragged to the center and merge with the large
galaxy. All of these processes alter the predicted mass function, but
to follow them all requires difficult-to-execute and physically
uncertain simulations.  From the point of view of observations, the
ram pressure stripping removes any existing neutral gas disks, the
component of dwarf galaxies known to extend furthest out into the dark
matter halo (\citealt{stoehr02a}). With only the stars, it is
difficult to probe any dark halo that might still surround a Local
Group dwarf.

On the other hand, isolated dwarf galaxies are possibly much simpler
systems. Without a large galaxy nearby, the physical processes
described in the previous paragraph cannot occur, simplifying the
prediction of their mass function. In addition, as it happens,
isolated dwarf galaxies essentially always have intact HI disks
(\citealt{geha06b}), allowing us to probe their masses out to large
radii with relative ease (\citealt{swaters02a}). The disadvantage of
isolated dwarf galaxies is that they are far away, making them
difficult to find at the lowest luminosities.

This paper represents a first step at examining the low mass end of
the mass function for isolated galaxies.  We use a sample of dwarf
galaxies selected from the Sloan Digital Sky Survey (SDSS;
\citealt{york00a}). From the
SDSS, we evaluate their number densities, and we estimate their
maximum circular velocity using single-beam HI profiles taken using
the Arecibo Observatory and the Green Bank Telescope.  We compare
these observations to theoretical predictions for the number density
of isolated halos of the same circular velocity.  By comparing
circular velocities, we avoid the considerable theoretical and
observational uncertainties in determining the total mass of halos and
galaxies.  While this approach represents a beginning, we discuss in
the conclusions how we can improve our observational estimates of the
masses and number densities of these objects, and also find lower mass
galaxies. With a much larger sample of isolated low mass galaxies, our
technique can provide a stringent and unique constraint on the power
spectrum at small scales.

In Section \ref{observations}, we describe the optical and radio
observations our results are based on.  In Section \ref{tullyfisher},
we describe the Tully-Fisher relationship for isolated galaxies. In
Section \ref{theory}, we describe our theoretical models.  In Section
\ref{comparison}, we compare the observations to the theory.  In
Section \ref{robust}, we explore the robustness of our results to our
definition of ``isolation'' in this context. In Section
\ref{fractions}, we examine the relationships among the baryonic,
stellar, and total mass estimates in our sample. In Section
\ref{future}, we describe how these results might be improved upon in
the future. Finally, in Section \ref{summary}, we summarize our
results.

For determining luminosity distances and other derived parameters from
observations, we have assumed cosmological parameters $\Omega_0 =
0.3$, $\Omega_\Lambda = 0.7$, and $H_0 = 100$ $h$ km s$^{-1}$
Mpc$^{-1}$. Where necessary, we have used $h=0.7$; otherwise, we have
left the dependence on $h$ explicit. All magnitudes in this paper are
$K$-corrected to rest-frame bandpasses using the method of
\citet{blanton03b} and {\tt kcorrect} {\tt v4\_1\_4}, unless otherwise
specified.  Because of the small range of look-back times in our
sample (a maximum of around 700 Myr), we do not evolution-correct any
of our magnitudes.

\section{ Observations}
\label{observations}

\subsection{ Sloan Digital Sky Survey }

To evaluate the luminosity function of dwarf galaxies, we use a
modified version of the SDSS spectroscopic catalog.
\citet{blanton04b} describe our sample, which is a subsample of the
New York University Value-Added Galaxy Catalog
\citep[NYU-VAGC;][]{blanton05a}. We have updated that catalog from
SDSS Data Release 2 to Data Release 4
\citep[DR4;][]{adelman06a}.  The \citet{blanton04b} catalog represents
a significant improvement over na\"ively selecting galaxies from the
SDSS catalog, which is not optimized for nearby, low surface
brightness galaxies.  It is selected with an optical flux limit of
$m_r \sim 17.8$.  For each galaxy, the catalog provides the SDSS
redshift, emission line measurements, multi-band photometry,
structural measurements and environment estimates (for more catalog
details see \citealt{blanton04b}).  Distances are estimated based on a
model of the local velocity field \citep{willick97a}.  Distance errors
have been folded into error estimates of all distance-dependent
quantities such as absolute magnitude and HI mass.

\begin{figure}
\figurenum{\fignum}
\plotone{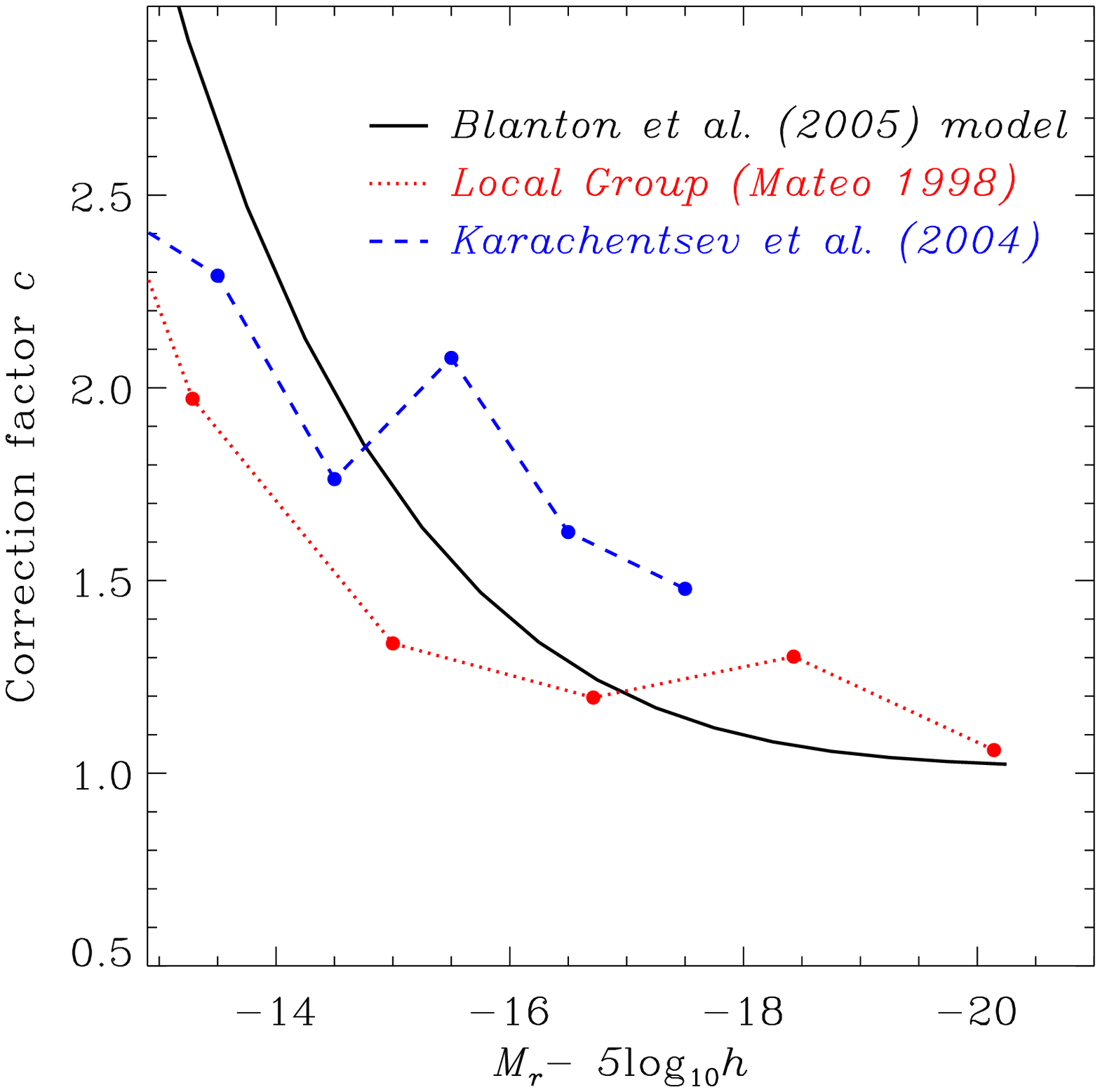}
\caption{\label{stupid_sb} Estimates of necessary correction factors
	$c$ for surface brightness incompleteness as a function of absolute
	magnitude. The solid line is the estimate from \citet{blanton04b}
	using a simple model of surface brightness as a function of absolute
	magnitude and determinations of the completeness of SDSS as a
	function of surface brightness.  The dashed line uses the
	distribution of surface brightnesses as a function of absolute
	magnitude from the local catalog of \citet{karachentsev04a}. The
	dotted line uses the distribution from the Local Group according to
	\citet{mateo98a}. }
\end{figure}

For our purposes, this catalog suffers from one major selection
effect, due to the difficulty of detecting low surface brightness
galaxies in the optical.  As shown in \citet{blanton04b}, the
completeness as a function of half-light surface brightness drops
below 50\% at $\mu_{50, r} \sim 23.5$ mag arcsec$^{-2}$.
\citet{blanton04b} present a simple model for the effect that surface
brightness has on the completeness, which assumes a log-normal surface
brightness distribution with a mean that decreases as luminosity
decreases. Figure \ref{stupid_sb} shows what this model implies about
the missing fraction of galaxies as a function of luminosity in our
sample, in terms of the correction factor $c$ we must apply to recover
the ``correct'' number density at each luminosity. For comparison, we
also show the equivalent factor for dwarf galaxies in the Local Group
\citep{mateo98a} given the SDSS surface brightness cutoff, and 
in the local catalog of \citet{karachentsev04a}. Note that the
\citet{karachentsev04a} curve is complicated by the fact that the
quantities in the catalog (Holmberg radius and flux) are not enough to
infer a half-light surface brightness (the terms in which we have
calculated the completeness) uniquely in general, even if we assume an
exponential profile. The \citet{mateo98a} and
\citet{karachentsev04a} catalogs bracket our correction factor of
$1.7$ at $M_r - 5\log_{10} h \sim 15$, indicating it is roughly
correct to within 20\%. Therefore, we use the model of
\citet{blanton04b} to correct the luminosity functions we present here
--- without any {\it a posteriori} corrections.  Nevertheless, the
uncertainty of how many galaxies we are missing due to this effect is
the most worrying one affecting our results.

\subsection{ Environments of galaxies }

As argued above, there are advantages to finding {\it isolated}
galaxies with which to test the circular velocity and mass functions.
We cannot rely on the SDSS alone to determine whether a galaxy is
isolated, for several reasons.  First, the angular distances between
nearest neighbor galaxies can be large for this nearby sample--- for
example, searching a 1\,Mpc region around a galaxy 30\,Mpc away
corresponds to 2 degrees on the sky.  Many of our dwarf galaxies are
on the SDSS Southern stripes, which are only 2.5 degrees wide.  In
addition, because the SDSS reduction software is not optimized for
large, extended objects and fails to process them correctly, the SDSS
catalog does not contain many of the bright galaxies within 30\,Mpc.
Thus, to calculate the environments of our dwarf galaxy sample, we
need a supplemental catalog that extends beyond the SDSS area and
contains the brightest galaxies.

Both of these considerations drive us to use the The Third Reference
Catalog of Galaxies (RC3; \citealt{devaucouleurs91a}), which is a
nearly complete catalog of nearby galaxies. To determine environments
for our dwarf galaxies, we must determine the distance of each to its
nearest ``luminous'' neighbor. In this context, we define galaxies as
luminous when $M_r - 5\log_{10} h < -19$, corresponding to circular
velocities of $V_c > 140$\kms\ for galaxies on the Tully-Fisher
relationship (see Section \ref{tullyfisher}).  From the $B$ and $V$
photometry listed in RC3, we infer $M_r$ for each galaxy.  For
galaxies which have the relevant entries listed, we call galaxies
luminous if $M_r -5 \log_{10} h< -19$. For galaxies which do not have
the relevant entries, but do have HI data listed, we call them
luminous if $W_{20} > 300$ km s$^{-1}$ (as described in
\S\,\ref{radio}, $W_{20}$ is twice the maximum circular velocity of
the HI gas).  Finally, there are some galaxies with neither HI data
nor optical photometry listed in RC3. For this small set, we extract
the ``magnitude'' from NED (which empirically is very similar to the
$B$ band RC3 magnitude for galaxies which have both) and guess $M_r$
based on that magnitude. We call these galaxies luminous if $M_r -
5\log_{10}h<-19$.  Additionally, we update the coordinates in RC3
using the NASA Extragalactic Database (NED) coordinates for each of
the catalog objects.  This set of bright galaxies is not perfectly
uniform, but is suitable for our purposes.

We combine the SDSS galaxies with $M_r - 5\log_{10} h< -19$ with the
RC3 luminous galaxy catalog (removing repeats between the two).  Then,
we determine the nearest neighbor distance by asking whether there is
a luminous RC3 or SDSS neighbor within 2 $h^{-1}$ Mpc and 400\kms\ in
redshift for each galaxy in our sample.  In this paper, we will
generally refer to isolated galaxies as those with no such neighbor
within $r_{\mathrm{lim}} = 1.0$ $h^{-1}$ Mpc. However, in order to
test the robustness of our results, we will vary our procedure below
by using alternate limits of $M_r - 5\log_{10} h< -18$ (a ``fainter
tracer sample'') and $M_r - 5\log_{10} h< -20$ (a ``brighter tracer
sample''). We will also vary the $r_P$ limit, as we describe
explicitly below.

We can calculate the number density of galaxies by weighting with the
$1/V_{\mathrm{max}}$ values that \citet{blanton05a} describes. Of
course, some of the volume actually is not available (e.g. in a
cluster a galaxy will never appear isolated) and in principle we might
need to account for this fact.  To do so, we can measure the
``isolated fraction,'' the fraction of the volume of our sample which
is isolated from luminous galaxies, by randomly placing points within
the volume and testing their environment.  This fraction is 0.95, and
varies by less than 0.05 when we vary our definition of ``isolated''
as noted above, minor compared to our other uncertainties. Thus, we
simply ignore this correction, and calculate the number density of
isolated galaxies over the entire volume.

\begin{figure}
\figurenum{\fignum}
\plotone{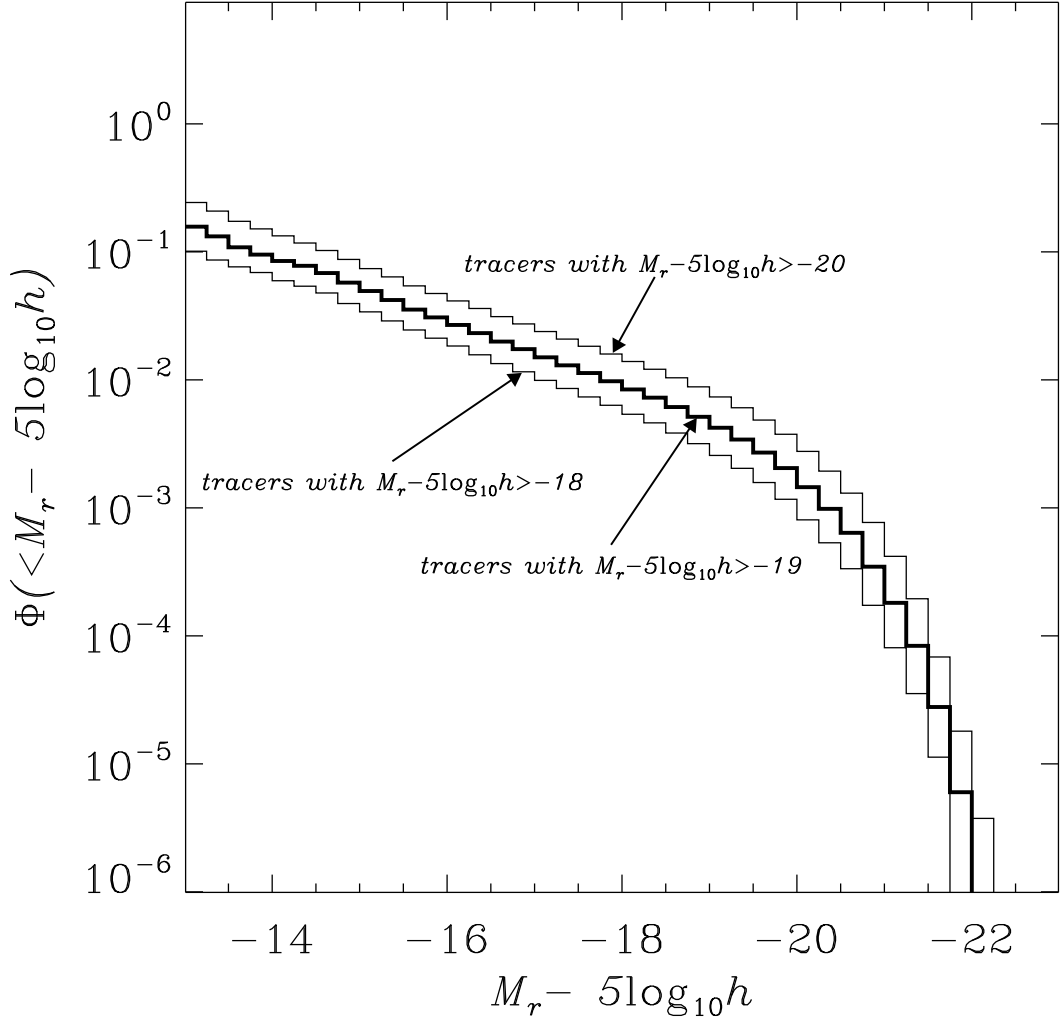}
\caption{\label{isovf_clf} Cumulative luminosity function of isolated
	galaxies, as defined in the text. The meaning of ``isolated''
	depends on the tracer galaxies used. The thick histogram represents
	the median relationship using tracer galaxies with $M_r - 5\log_{10}
	h < -19$.  The thin histogram represents a change of 1 magnitude in
	the absolute magnitude limit used for the tracer galaxies; the upper
	histogram uses brighter galaxies, and the lower histogram uses
	fainter galaxies. }
\end{figure}

The thick line Figure \ref{isovf_clf} shows the cumulative number
density of such isolated galaxies as a function of absolute magnitude,
corrected for surface brightness incompleteness.  This statistic
depends on the definition of ``isolated'' that we use; the two thin
lines use the fainter and brighter tracer samples described above. As
this comparison shows, a change in magnitude of the tracer by about 1
mag is equivalent to a 20\% change in the cumulative number density at
$M_r - 5\log_{10} h \sim -15$. 

Taken together these data show that the number density of isolated
galaxies with $M_r-5\log_{10} h <\mrmean$ is $(6.83 \pm 2.3)\times
10^{-2}$ $h^3$ Mpc$^{-3}$, with the uncertainty dominated by the
definition of ``isolated'' and the fraction of galaxies missing due to
surface brightness effects.

\subsection{ Radio observations at 21cm }
\label{radio}

\citet{geha06b} present the radio observations of dwarf galaxies
selected from the SDSS sample described above.  The data were obtained
on the Green Bank 100-m Telescope (GBT) and at Arecibo
Observatory. Each observation had a velocity resolution better than
about 3 km s$^{-1}$. The optical half-light radii of the dwarf
galaxies in our sample are typically $\sim 8''$, and should be
completely contained with the radio beamsize of $3'$ and $9'$ for
Arecibo and GBT, respectively.  Here we will restrict ourselves to
galaxies with detected HI (which \citealt{geha06b} showed dominated
the isolated galaxy population) and with $b/a< 0.5$ to minimize
inclination effects.

We compute the 20\% HI line-width (W20) by finding the peak HI flux
within 150\kms\ of the optical radial velocity of each galaxy and
computing the difference between the nearest points having 20\% of the
peak flux.  The integrated HI flux is calculated by expanding the W20
values by 20\kms\ on each side and integrating the flux in this
region.  Errors bars on the line widths and integrated fluxes were
computed using a Monte Carlo bootstrap method: noise was added to the
stacked one-dimensional radio spectra (based on the observed variance
in the baseline) and the observed quantities remeasured.  We
calculated error bars on the line-width and integrated flux from the
scatter in the mean quantities recovered from Monte Carlo simulations.

As \citet{geha06b} showed, the dwarf galaxies in our sample have a
significant rotation component. We derive the maximum rotation speeds
as follows.  We correct the observed HI line-widths for line
broadening due to turbulent velocity dispersion and inclination using
the formula first proposed by \citet{bottinelli83a}:
\begin{equation}
\label{bottinelli}
V_{\mathrm{max}} = \frac{\mathrm{W20}-\mathrm{W20}_{\it t}}{\sin{\it
		i}},
\end{equation}
where W20 is the observed HI line-width, W20$_t$ is the turbulent
velocity correction term and $i$ is the inclination angle inferred
from the optical images.  We confirm the validity of a linear
turbulence correction by modeling the integrated velocity profiles of
simulated galaxies constructed from \citet{hernquist93a} model disk
galaxies.  For nearby dwarf galaxies with rotation velocities similar
to our sample, \citet{begum06a} have measured a velocity dispersion in
the gas component of $\sigma_{\rm los} = 8$\kms\ from 2D velocity
maps.  Using the Begum et al.~value, this results in a turbulence
correction of W20$_t = 16$\kms, which we use here.  Altering this
turbulence correction (say to 25 \kms) does not change our results
below.

\begin{figure}
\figurenum{\fignum}
\plotone{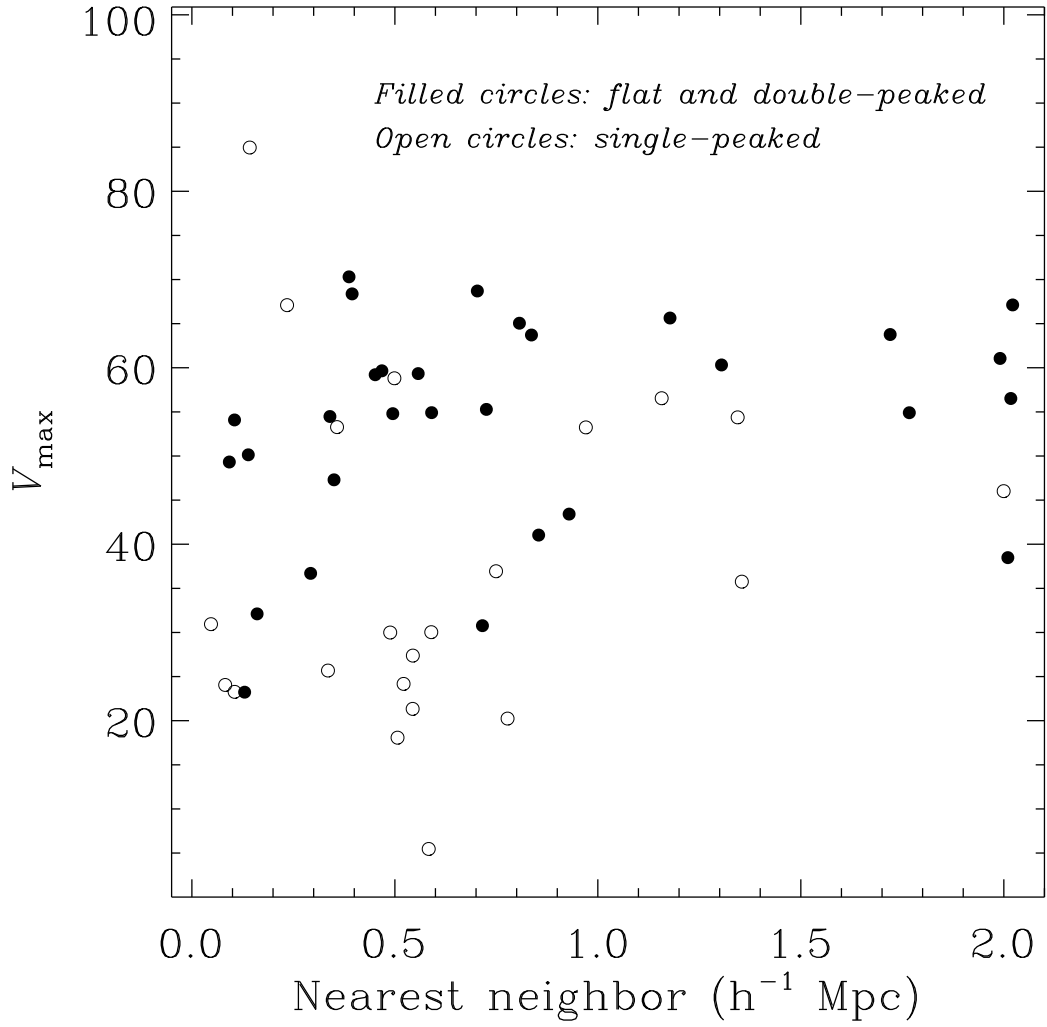}
\caption{\label{w20_vs_rp} Estimate of $V_{\mathrm{max}}$ for dwarf
	galaxies in our sample as a function of distance to the nearest
	luminous neighbor galaxy. Filled symbols are galaxies with flat or
	double-peaked HI profiles, open symbols are galaxies with
	single-peaked HI profiles. }
\end{figure}

Figure \ref{w20_vs_rp} shows our best estimate of $V_{\mathrm{max}}$
for our dwarf galaxies (with $b/a<0.5$) as a function of the distance
to the nearest luminous luminous galaxy.  Clearly there is a strong
relationship between $V_{\mathrm{max}}$ and projected separation.
Within 1 $h^{-1}$ Mpc of luminous galaxies there is a population of
objects with $V_{\mathrm{max}} < 40$ km s$^{-1}$, that is much rarer
at large separations.  Our galaxies are selected to be distributed
roughly evenly in the range $-13.5 > M_r - 5 \log_{10} h > -15.5$,
independent of environment.  Thus, at these luminosities the
``forward'' Tully-Fisher relationship --- the circular velocity at a
fixed luminosity --- appears to be a strong function of nearest
neighbor distance.  This result does not imply that no low circular
velocity galaxies exist in the field, simply that any such galaxies
are too low luminosity to make it into our sample.

As \citet{geha06b} and others have found, dwarf galaxies near a
luminous neighbor also tend to be red and gas poor.  Taken all
together, these results suggest that some important physical effects
are shaping the gas content, star-formation histories and inferred
dynamics of dwarf satellite galaxies relative to isolated dwarfs.  For
this reason, we choose to concentrate our attention here on the
isolated galaxies, whose $V_{\mathrm{max}}$ values have a smaller
dispersion, and whose properties in general we expect to be less
altered since formation, relative to those dwarfs perturbed by a
massive neighbor.

Although it is immaterial to our analysis below, it is interesting to
ask what physical effects are causing the trend in Figure
\ref{w20_vs_rp}. We can think of three explanations. First, for dwarfs
in the vicinity of luminous galaxies, ram pressure stripping could
remove gas at the largest galactocentric radii first, reducing the
maximum circular velocity traced by HI emission. Second, tidal
stripping could reduce the total mass and thus the circular
velocities.  Third, interaction-triggered star-formation could either
raise the luminosities of satellite galaxies relative to isolated
galaxies, bringing lower mass systems into our sample if they are near
bright galaxies, or speed up star-formation and use up the gas in the
outer disk of the galaxy. To help investigate this question, in Figure
\ref{w20_vs_rp} we have distinguished between single-peaked HI
profiles (open circles) and double-peaked or flat-topped profiles
(filled circles), as classified by \citealt{geha06b}).  Because the
dwarfs with low $V_{\mathrm{max}}$ are predominantly single-peaked
profiles (and we have enough resolution to see double-peaked profiles
if they existed), we favor the explanation that gas has been stripped
from the outsides. However, a definitive conclusion awaits a
comprehensive analysis, including more detailed dynamics of these
galaxies.

\section{ Tully-Fisher relation for isolated galaxies}
\label{tullyfisher}

Here we give an estimate of the Tully-Fisher relationship for isolated
galaxies. We base our estimate on the sample described in the previous
section for low luminosities, plus the recently published samples of
\citet{pizagno06a} and \citet{springob05a} for higher luminosities.

\citet{pizagno06a} presented a Tully-Fisher survey of galaxies found
in the SDSS, using follow-up H$\alpha$ rotation curves. They have
performed model fits to the disk components of these galaxies to
inclination-correct their maximum circular velocities.  Unlike
previous Tully-Fisher samples (e.g., \citealt{courteau97a}), which
were selected to be very homogeneous sets of galaxies, the
\citet{pizagno06a} sample spans a large range of galaxy types and
colors, resulting in a somewhat larger scatter in the Tully-Fisher
relationship than found in other studies.  For each of their galaxies
we have determined its environment in the same manner as for our low
luminosity sample, and we only consider isolated galaxies here.
However, as \citet{pizagno06a} show, and we have confirmed using our
own measurements of environment, there is very little dependence of
the Tully-Fisher relation on environment at high luminosities. As with
the low luminosity sample, we restrict ourselves to galaxies with $b/a
< 0.5$, leaving 35 galaxes from \citet{pizagno06a}.

\citet{springob05a} have compiled HI spectra from archival data sets
for around 9000 galaxies in the local Universe, observed originally
with Arecibo, the 91m and 42m Green Bank telescopes, the Nan\c{c}ay
telescope, and the Effelsberg 100m telescope. They have homogeneously
analyzed these spectra, measuring their widths and HI fluxes.  All
their galaxies have optical data associated with them, including a
measure of their axis ratio ($b/a$). We use the turbulence and
inclination corrections of Equation (\ref{bottinelli}) to convert the
$W_{20}$ values to $V_{\mathrm{max}}$. In addition, in order to get
luminosities and environments for these galaxies, we match this list
to the full low-redshift catalog from SDSS DR4 (just as for our dwarf
sample).  Finally, we restrict this sample to the isolated galaxies
with $b/a < 0.5$, leaving 85 galaxies from \citet{springob05a}.

\begin{figure}
\figurenum{\fignum}
\plotone{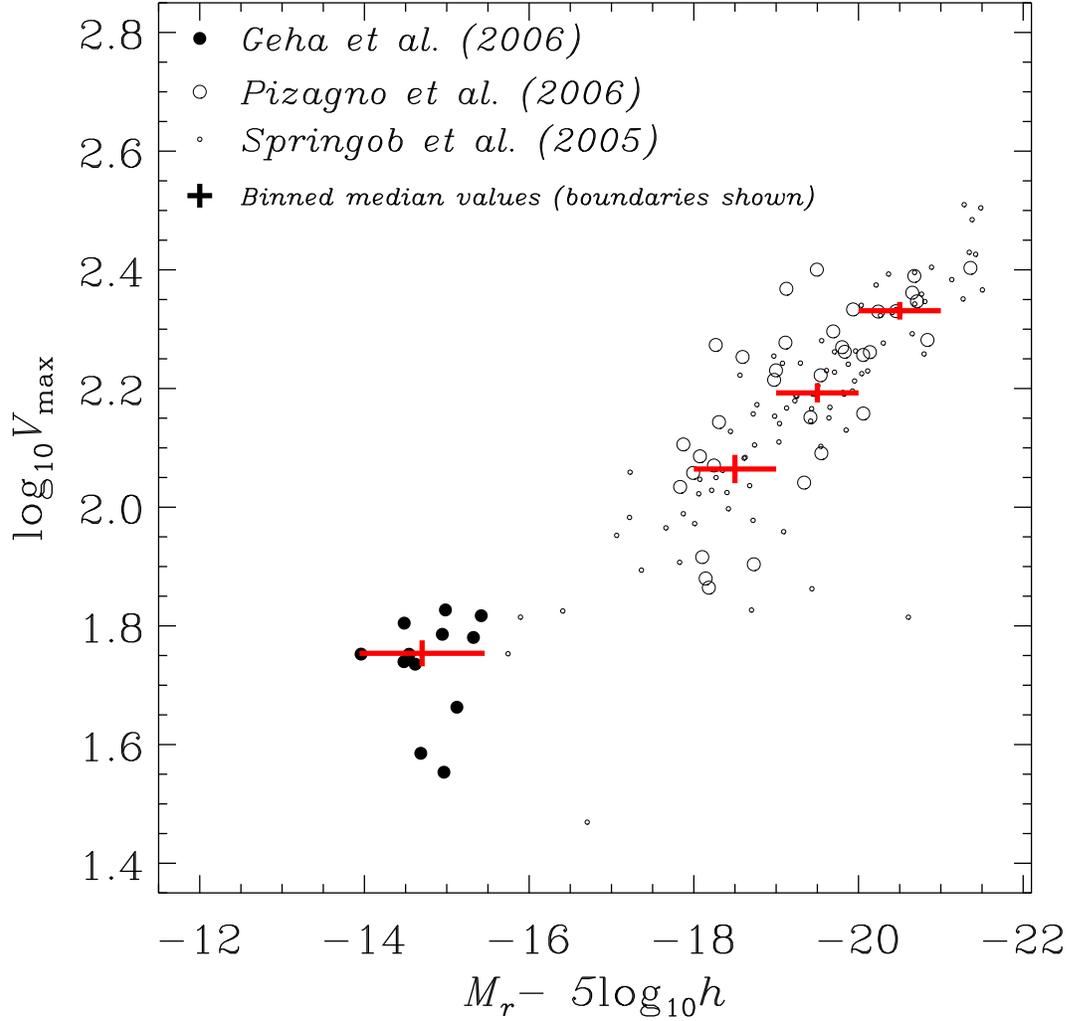}
\caption{\label{isotf_dataonly} Tully-Fisher relation for isolated
	galaxies.  Large open symbols are from the optical rotation curves
	of \citet{pizagno06a}, small open symbols are from the compilation
	of \citet{springob05a}, and filled symbols are from the HI
	linewidths of \citet{geha06a}. Crosses show the median
	$V_{\mathrm{max}}$ values in several bins; the bin centers and
	median values are listed in Table \ref{vmaxchoice}.}
\end{figure}

Figure \ref{isotf_dataonly} shows the Tully-Fisher relationship for
all three samples of galaxies. The low luminosity points are from our
isolated sample.  The crosses in Figure \ref{isotf_dataonly} show the median
$V_{\mathrm{max}}$ values in several bins centered on $M_r -
5\log_{10} h = -14.7, -18.5$, $-19.5$, and $-20.5$, and 1 magnitude in
width. Table \ref{vmaxchoice} lists the median velocity in each bin
(and several other bins needed below).

\section{ Theory }
\label{theory}

In order to find an appropriate theoretical comparison, we use the
$N$-body, pure dark matter simulations of \citet{kravtsov04b}.  They
simulated a cubic box $80$ $h^{-1}$ Mpc on a side, with 512$^{3}$
particles, each around $3.2\times 10^8$ $h^{-1}$ $M_\odot$ in
mass. The world model and transfer function for the simulation
correspond to $\Omega_{m} = 0.24$, $\Omega_\Lambda =0.76$, $\Omega_b =
0$, $h=0.73$, $n=0.95$, and $\sigma_8 =0.75$. Although this cosmology
is not {\it precisely} the current best-fit model, we describe
below how we implement small corrections to account for this fact. We used
the outputs from $z=0$. 

Dark matter halos were identified using the method described in
\citet{kravtsov04b}. They calculated the maximum circular velocity 
of each halo by determining the enclosed mass $M(<R)$ as a function of
radius, using the spherically symmetric approximation
$v=\sqrt{GM(<R)/R}$, and determining the peak in the rotation
curve. Strictly speaking, this $V_{\mathrm{max}}$ value is only
comparable to our observed galaxies if their 21cm emission probes this
peak. However, because most of our galaxies show flat-topped or
double-peaked HI profiles, we are probably close to satisfying this
condition.

\begin{figure}
\figurenum{\fignum}
\plotone{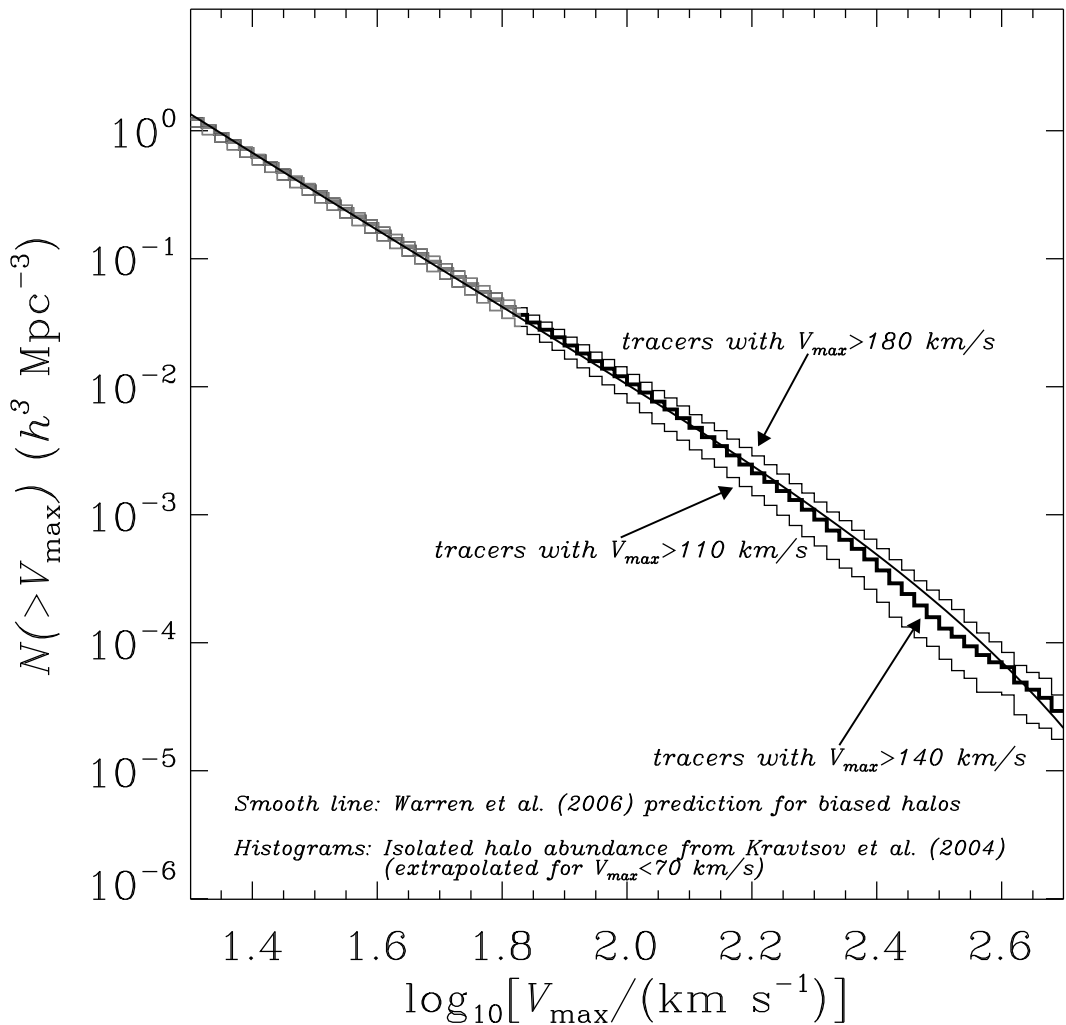}
\caption{\label{isovf_cvf} Cumulative $V_{\mathrm{max}}$ function of
	isolated halos from $N$-body simulations, for various choices of
	tracers.  The thick histogram is the abundance of halos in the
	simulation subject to the isolation criterion described in the text,
	based on the projected distance and redshift different relative to
	the nearest, other ``tracer'' halo, for tracers with
	$V_{\mathrm{max}}>140$ km s$^{-1}$. The thin histograms explore the
	effect of trying different tracer populations, as labeled.  The smooth
	line is an approximation based on the mass functions of
	\citet{warren06a}, described more fully in the text, using a
	relative large-scale bias of $\delta=-0.4$ for the halos.}
\end{figure}

To define ``isolated'' we use tracer halos with $V_{\mathrm{max}}$
values corresponding to the Tully-Fisher results listed in Table
\ref{vmaxchoice}.  We project the distribution of halos in a random
direction, and perturb the ``redshift-space'' positions of the halos
to account for their peculiar velocities.  Then we define ``isolated''
in the simulation using the same geometrical considerations used in
the observations ($|\Delta v| < 400$ km s$^{-1}$ and $r_p < 1$
$h^{-1}$ Mpc) but relative to the tracer halo population.  The
velocity function of these isolated halos is shown in Figure
\ref{isovf_cvf} as the histograms.  Down to 70 km s$^{-1}$ each
histogram comes from the simulation, but below that we extrapolate the
velocity function as a power law (roughly
$\Phi(>V_{\mathrm{max}})\propto V_{\mathrm{max}}^{-2.7}$). The three
histograms, as labeled, correspond to tracers with
$V_{\mathrm{max}}>110$ km s$^{-1}$ (corresponding to galaxies with
$M_r -5 \log_{10} h < -18$), $V_{\mathrm{max}}>140$ km s$^{-1}$
(corresponding to galaxies with $M_r -5 \log_{10} h < -19$), and
$V_{\mathrm{max}}>180$ km s$^{-1}$ (corresponding to galaxies with
$M_r -5 \log_{10} h < -20$). For the bulk of this paper, we will be
concerned with the central class, shown as the thick histogram, but
will use the other results to quantify how much our results depend on
the choice of tracer population.

Because of the complex geometrical definition of ``isolated,'' it is
useful to have this $N$-body estimate of the mass function.  However,
the cosmology used for the simulation does not correspond precisely to
the current best fit cosmology (e.g. \citealt{tegmark06a}); in
particular it does not include the effects of baryons on the initial
power spectrum.  We adjust for this difference by using the excursion
set formalism and the transfer functions of \citet{eisenstein98a},
along with the mass function approximation from
\citet{warren06a}. These methods are able to predict the mass function
for any hierarchical cosmology, as a function of redshift and of
large-scale environment. We use a particular implementation provided
by Andreas Berlind (private communication). These methods yield the
mass function of galaxies, which we convert into a circular velocity
function using the methods of \citet{bullock01b}, using $M_\ast =
1.5\times 10^{12}$ $h^{-1}$ $M_\odot$ and $\Omega_m = 0.24$.

First, we need to evaluate what large-scale environment our definition
of ``isolated'' corresponds to. The smooth line in Figure
\ref{isovf_cvf} is the prediction for the mass function of halos in
large-scale underdensities of $\delta=-0.4$, for the cosmology used in
\citet{kravtsov04b}. From this agreement, we conclude that in the
excursion set mass functions, $\delta=-0.4$ is the underdensity that
is most comparable to our isolation criterion.

\begin{figure}
\figurenum{\fignum}
\plotone{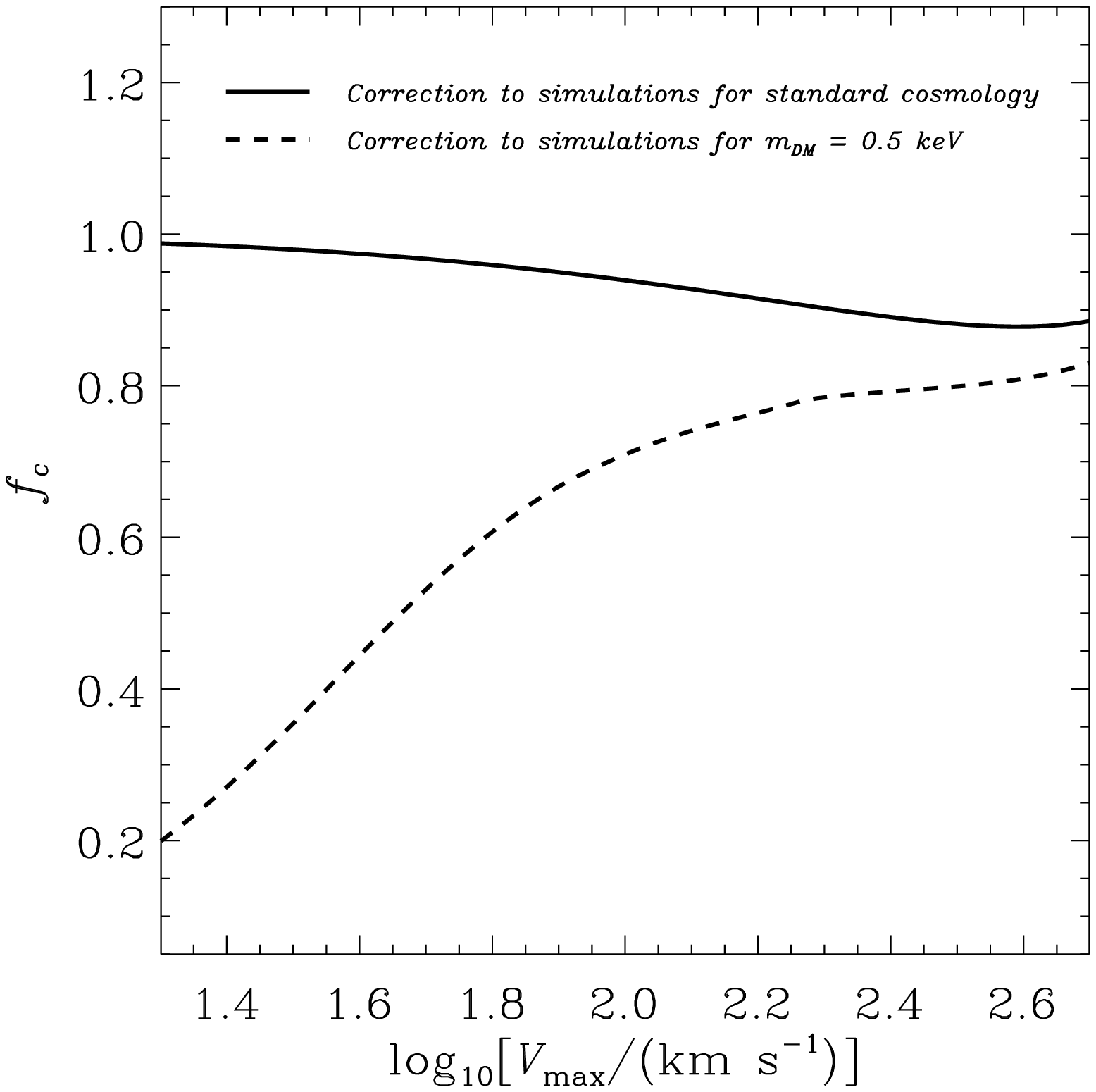}
\caption{\label{isovf_correct} Correction to the mass function between
	the cosmology of the $N$-body simulations of \citet{kravtsov04b} and
	the cosmological models we want to test. The solid line is the
	correction to the standard cosmology of \citet{tegmark06a} based on
	WMAP and SDSS large-scale data.  The dashed line is the correction
	to that same cosmology, but with a light dark matter particle
	($m_{\mathrm{DM}} = \mdmlimit$ keV). Corrections are based on the ratio
	between the appropriate cosmologies in the approximations of
	\citet{warren06a}.}
\end{figure}

Second, in order to adjust the results of \citet{kravtsov04b} for
cosmology, we evaluate the ratio $f_c$ between the velocity function
for the cosmology of \citet{kravtsov04b} (listed above) and the best
fit of \citet{tegmark06a} (the only differences are that in the latter
$\sigma_8=0.76$ and $\Omega_b h^2 = 0.022$).  Figure
\ref{isovf_correct} shows this ratio as a function of maximum circular
velocity. Over the range we will use here, this correction is never
more than about 10\%. In order to compare our results from the
\citet{kravtsov04b} simulations to observations, we first apply this
correction factor to the predictions from the simulations.

In addition, we have incorporated a ``warm dark matter'' version of
the \citet{tegmark06a} cosmology that is identical in its large-scale
structure, but includes a light dark matter particle, with $m_{DM} =
\mdmlimit$ keV.  The lightness of this particle increases its free streaming
length, which smooths fluctuations on small scales.  Here we apply the
adjustment required to the transfer function as outlined by
\citet{abazajian06a}. We simply input this new transfer function into
the excursion set calculation of the mass function.  Of course, on the
scales comparable to the free-streaming length, the collapse of
structure will cease to be hierarchical, as described by
\citet{bode01a}. Thus, almost by definition this prediction will be
incorrect; however, we use it as a rough approximation to a more
correct prediction which might be available in the future.  Figure
\ref{isovf_correct} shows the ``correction'' as a function of
$V_{\mathrm{max}}$ as the dashed line. In order to predict the warm
dark matter case, we apply this correction factor to the simulations.

Figure \ref{isovf_cvf_corr} shows the resulting $V_{\mathrm{max}}$
functions in the cold dark matter and warm dark matter cases. In the
next section, we describe how we compare these theoretical predictions
to the observations.

\section{ Comparing theory to observations}
\label{comparison}

The simplest possible comparison we can make between simulations and
the observations is to try to put some observed points on the velocity
function. After all, we have measured the number density of galaxies
as a function of luminosity, and from the Tully-Fisher measurements,
we know the relationship between luminosity and the circular
velocities. Given the luminosity function of Figure \ref{isovf_clf}
(using the tracers with $M_r -5\log_{10}h < -19$), plus the
relationship between luminosity and $V_{\mathrm{max}}$ from Figure
\ref{isotf_dataonly}, we also plot the observed number density as a
function of $V_\mathrm{max}$ as the four points.  In this comparison,
there appear to be slight discrepancies between the cold dark matter
model and the observations, particular at the bright end. The warm
dark matter model is nearly as good a fit to the data but somewhat
underpredicts (at a bit more than 1$\sigma$) the number of isolated
low circular velocity galaxies.

However, this method of comparison is sensitive to bias related to
scatter in the relationship between luminosity and circular
velocity. A more robust comparison can be achieved as follows. For a
given predicted circular velocity function and a given observed
luminosity function, we can find the relationship between circular
velocity and luminosity that makes them consistent with one
another. We do so here by parameterizing it as a piecewise linear
relationship $M_r(V_{\mathrm{max}})$ between the circular velocity and
a mean absolute magnitude, with Gaussian scatter about that mean. We
vary the parameters of the piecewise linear relationship to fit the
luminosity function (using the Levenberg-Marquardt method implemented
in the IDL routine {\tt mpfit} distributed by Craig
B.~Markwardt). However, we fix the Gaussian scatter to have $\sigma_M
= 2.5$ for $V_{\mathrm{max}}\le 10$ km s$^{-1}$, to have $\sigma_M
=0.4$ for $V_{\mathrm{max}}\ge 100$ km s$^{-1}$, and to vary linearly
with circular velocity in between.

Figure \ref{isotf} shows the conditional distribution of
$V_{\mathrm{max}}$ as a function of $M_r$, given our best fit
relationship (using tracers in the simulation with
$V_{\mathrm{max}}>155$ km s$^{-1}$ and corresponding tracers in the
observations with $M_r - 5\log_{10} h < -19$). The lines are the
quartiles of the distribution. The overplotted points are the data
from Figure \ref{isotf_dataonly}. Clearly the median values (shown as
the boxes) agree rather well with the predictions.  This constitutes a
confirmation that the Tully-Fisher relationship and the luminosity
function together are consistent with cold dark matter predictions.

How well can these data exclude alternate scenarios? We explore this
question by considering the warm dark matter model described above,
with a dark matter particle mass of $\mdmlimit$ keV. We perform the same
procedure as described above and obtain the predictions shown in
Figure \ref{isotf_warm}. Here, the observed circular velocities of low
mass galaxies tend to be higher than predicted, but not by significant
amounts. The median is about 4$\sigma$ away from the predicted median,
which given the systematic uncertainties here we regard as a marginal
exclusion of this model.

\section{ Robustness relative to our definition of ``isolated''} 
\label{robust}

Here we examine the robustness of our results to our definition of
``isolation,'' and how we relate isolation in the observations to
isolation in the simulations.  There are a number of arbitrary
decisions we have made here to define ``isolated'' in the
observations. In particular, we chose a certain projected distance
$r_P$ from galaxies of a certain absolute magnitude
$M_{r,\mathrm{bright}}$ (or brighter). We must examine the sensitivity
of our results to these arbitrary choices.  In addition, we have also
had to define ``isolated'' in the theoretical predictions. To do so,
we have had to relate the absolute magnitude $M_r$ used above to a
$V_{\mathrm{max}}$ of the halos in the simulation. Of course, we do
not know the exact correspondence, and so we need to examine how our
results depend on errors in our estimate of it.

Figure \ref{isotf_all} examines the sensitivity of our results to both
sets of decisions. The top panels show the ratio of the predicted
$\Phi_{\mathrm{iso}}(>V_{\mathrm{max}})$ at $V_{\mathrm{max}} =
\vmaxmean$ km s$^{-1}$, to the observed $\Phi_{\mathrm{iso}}(<M_r)$ at
$M_r - 5\log_{10} h = \mrmean$. The bottom panels show another
comparison of the observations to the theory: the ratio of the
observed $V_{\mathrm{max}}=\vmaxmean$ km s$^{-1}$ at $M_r - 5\log_{10}
h = \mrmean$ to that predicted by equating the number density of halos
larger than a given $V_{\mathrm{max}}$ to the number density of
galaxies brighter than $M_r - 5\log_{10} h = \mrmean$. The left panels
correspond to our cold dark matter model, and the right panels
correspond to our warm dark matter model with $m_{DM} = \mdmlimit$ keV.

There are 27 points shown on each plot, each corresponding to a
different definition of ``isolated'' in the observations and the
theory.  First, we check three different choices of tracer sample
($M_{r,\mathrm{bright}}-5\log_{10}h = -18$, $-19$, and $-20$), as
shown by the rough horizontal position.  Second, for each of these
three choices of galaxy sample, we choose three different
$V_{\mathrm{max}}$ values with which to define our halo sample for
comparison. Our choices of predicted circular velocities for each
observed sample are listed in Table \ref{vmaxchoice}. In Figure
\ref{isotf_all}, the larger symbols correspond to higher circular
velocities.  Third, for each of those nine choices, we choose three
different choices of $r_P$ (0.7, 1.0, and 1.3 $h^{-1}$ Mpc), using the
same radius for theory and observation. These three cases are offset
from each other slightly in Figure \ref{isotf_all} (left to right,
respectively) for clarity.

From these results, it is clear that our systematic uncertainties are
(fractionally) about 0.3 in the comparison of number densities, and
0.1 in the comparison of circular velocities. The difference is at
least partly due to the approximate dependence
$\Phi(>V_{\mathrm{max}}) \propto V_{\mathrm{max}}^{-2.7}$. In
addition, we expect the comparison of circular velocities to be more
robust, since it depends less on the scatter in the relationship
between luminosity and circular velocity.  While for our sample, these
systematics are about equivalent to the systematics associated with
our surface brightness completeness selection, the systematics shown
in Figure \ref{isotf_all} illustrate what will ultimately be the most
difficult uncertainty to overcome, even when much more complete galaxy
samples are available.

\section{ Baryonic and stellar content of low mass galaxies }
\label{fractions}

In the previous section, we showed that the cold dark matter model
reasonably explains the number densities and circular velocities of
low luminosity galaxies. Assuming that the relationship between
circular velocity and total mass that cold dark matter theory predicts
is correct, we can now investigate the baryonic and stellar mass
content (relative to the total mass) of these low mass galaxies.  This
census of the matter in dwarf galaxies may help us understand their
creation and development over time.

To infer the total mass from $V_{\mathrm{max}}$ we use the methods of
\citet{bullock01b}, using $M_\ast = 1.5\times 10^{12}$ $h^{-1}$
$M_\odot$ and $\Omega_m = 0.24$.  These methods have been calibrated
down to masses of $10^{11}$ $h^{-1}$ $M_\odot$, or about $85$ km
s$^{-1}$, so our use of them for the lowest luminosity galaxies
represents an extrapolation of the current theoretical understanding.
At our typical $V_{\mathrm{max}} \sim 56$ km s$^{-1}$, the virial mass
of the halo determined by this method is $2.5\times 10^{10}$ $h^{-1}$
$M_\odot$.

To infer the stellar mass, we use the optical broadband SDSS data and
the methods of \citet{blanton06b}. Any method for inferring the total
stellar mass is sensitive to the initial mass function (IMF) of stars,
since the lowest mass stars (most of those below 0.5 $M_\odot$)
contribute almost no optical light but are a significant fraction of
the mass. The differences between different reasonable choices of IMF
can be up to 50\%. \citet{blanton06b} have chosen the
\citet{chabrier03a} IMF, and find stellar masses within about 30\% of
those found for the same galaxies by \citet{kauffmann03a}, using
spectroscopic techniques.  The median stellar mass for our isolated
dwarf galaxy sample is $2.2\times 10^7$ $h^{-2}$ $M_\odot$.

By ``baryonic mass,'' we mean here the sum of the stellar mass and
neutral gas content, which we take to be $M_b = M_\ast + 1.4
M_{\mathrm{HI}}$, where $M_{\mathrm{HI}}$ is the neutral hydrogen mass
inferred from 21cm observations and the factor $1.4$ accounts for
helium, molecular clouds and metals. The median baryonic mass so
defined is $2.5\times 10^8$ $h^{-2}$ $M_\odot$ --- much larger than
the stellar mass contribution.  Naturally, there may also be ionized
hydrogen in the galaxy, which we do not try to account for here. Note
for the data from \citet{geha06b} we do not try to correct for
self-absorption of HI, because little evidence for any inclination
dependence of the HI to stellar mass ratio is found in our
sample. However, \citet{springob05a} did make such corrections, which
can be up to 20\%.

\begin{figure}
\figurenum{\fignum}
\plotone{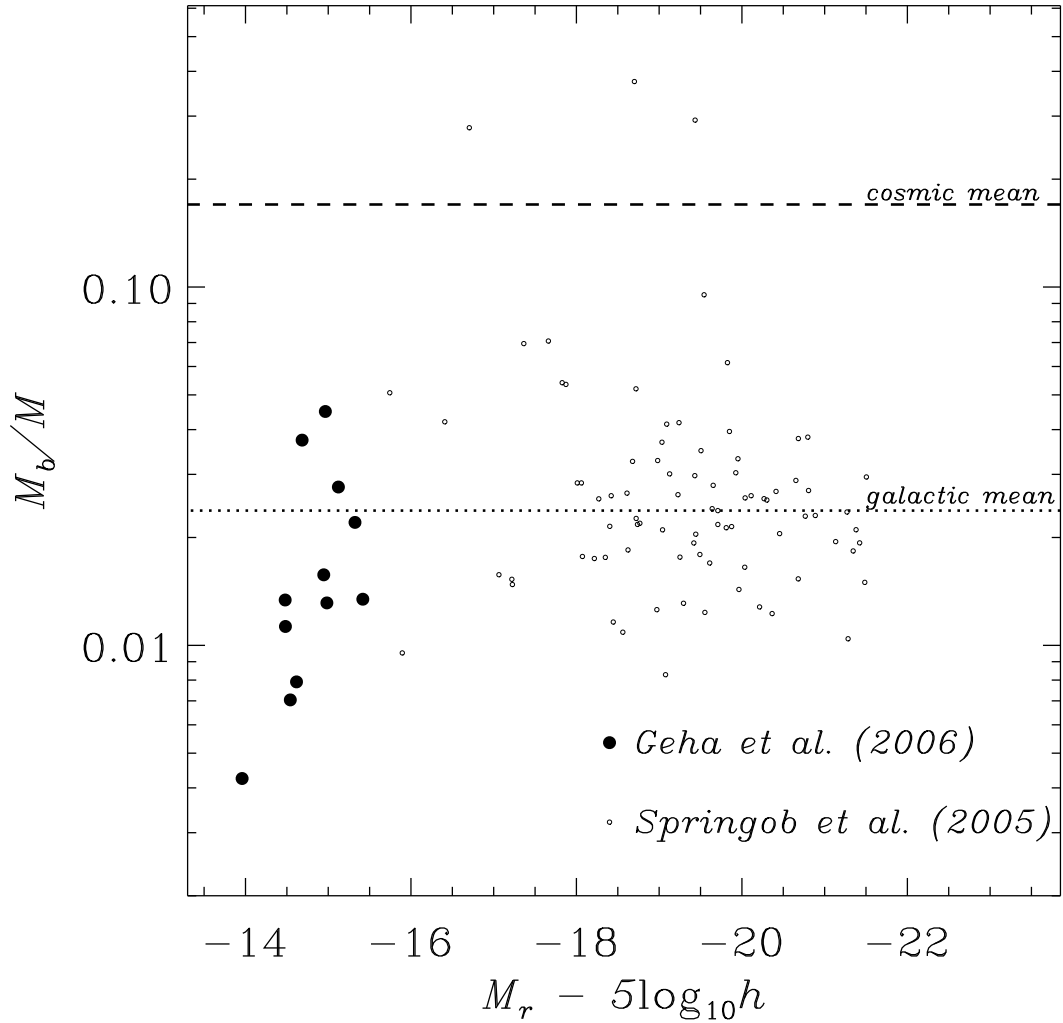}
\caption{\label{isotf_mtobm} ``Baryonic'' mass relative to total mass
	as a function of luminosity for the samples of isolated galaxies
	with $b/a<0.5$ in the two HI samples used here, as marked. Baryonic
	mass is defined as the stellar mass plus the neutral gas mass, as
	described in the text. The dashed line at $0.17$ is the cosmic mean
	based on cosmological measurements (\citealt{tegmark06a}). The
	dotted line at $0.025$ is the mean of the \citet{springob05a}
	measurements. }
\end{figure}

For the samples of isolated galaxies used in this paper, Figure
\ref{isotf_mtobm} shows the ratio of baryonic mass to total mass as a
function of $r$-band absolute magnitude (for $h=0.7$).  The dashed
line is the cosmic mean based on the results of \citet{tegmark06a}
($\Omega_b/\Omega_m = 0.17$) and the dotted line is the mean for the
sample of \citet{springob05a} taken alone. The mean of the low
luminosity galaxies from \citet{geha06b} is somewhat less than that of
higher luminosity galaxies, about 8\% of the cosmic value rather than
14\%. The difference in the treatment of self-absorption may account
for about half of this difference.

The baryonic fraction may continue to decrease at lower
masses. However, at $V_{\mathrm{max}} \sim 50$ km s$^{-1}$ isolated
low luminosity galaxies do not show much evidence that they have
expelled or ionized very much more of their cold gas than have their
more massive counterparts. This measurement supports detailed models
of the physics of gas blow-out and blow-way due to supernovae, which
predict a loss of only a few percent for galaxies in this mass range
\citep{maclow99a,ferrara00a,stinson07a}.  However, it disfavors models that
prevent star-formation in low mass galaxies through significant
baryonic mass loss \citep{dekel86a, cole00a, mori00a, bullock00a,
benson03a,dekel03a, tremonti04a, croton06a}. 
Of course, reasonable
modifications to those models in which feedback prevents the gas from
forming stars but keeps it mostly in neutral form and within the
galaxy disk are probably tenable.
Furthermore, although some 
investigators have invoked outflows to explain the mass-metallicity
relationship, \citet{dalcanton07a} have shown that
alternate models without significant outflow can explain the 
observations.  

We can also look at the relationship between the stellar mass of the
galaxies and their total mass. Figure \ref{isotf_mtosm} shows the
dynamical to stellar mass ratio as a function of stellar mass for the
three samples used here. This relationship shows a strong trend,
illustrating the strong dependence of star-formation efficiency on
mass, at least for disk galaxies. It is, of course, exactly this
dependence that causes the luminosity and stellar mass functions to be
shallow while the total mass function of galaxies is so steep, as
described in Section \ref{intro}.

\section{ Future directions }
\label{future}

The analysis of this paper, while consistent with the CDM model, puts
only mild constraints on alternative models (for example, a warm dark
matter particle as light as $\mdmlimit$ keV is barely ruled out).  How
can this analysis be improved in the future? Two paths are possible:
first, increasing the depth and completeness of our optical plus HI
sample; second, using upcoming blind HI surveys to push to
considerably lower masses.  Both paths require improving our
understanding of the theoretical predictions at the low mass end.

Our analysis of the current SDSS sample would be substantially
improved with more HI follow-up observations. Because \citet{geha06b}
were studying the general HI properties of dwarfs, we only targeted
about 12 systems edge-on enough and isolated enough to include in the
analysis of this paper. Our results could be put on a much firmer
footing with an increase in our follow up sample. In DR4 there are 64
galaxies with $b/a<0.5$, $M_r-5\log_{10}h>-15$ and $r_P>1$ $h^{-1}$
Mpc, and obviously there are still more in later releases.  Additional
follow-up or deeper HI sky surveys should fill this gap in the future.

The optical SDSS sample on which our analysis here is based will
increase by DR8, the final SDSS release, perhaps by a factor of
two. However, this will not substantially reduce the uncertainties in
the luminosity function, which are already dominated by the surface
brightness completeness correction (Figure
\ref{stupid_sb}). \citet{blanton04b} concluded, based on introducing
simulated galaxies into raw SDSS imaging data, that the SDSS
photometric pipeline (optimized for reasonably high surface brightness
galaxies around $z\sim 0.1$) was probably failing at a brighter
surface brightness limit than the data required, and that with a
differently optimized pipeline it might be possible to push the
surface brightness limits to 25 mag or more. Doing so would allow us
to probe magnitudes as faint as $M_r - 5\log_{10} h \sim -12$ over
cosmological volumes.  Another possibility would be to wait for
upcoming, deeper surveys such as the Dark Energy Survey (DES;
\citealt{wester05a}), which has first light in late 2010, Pan-STARRS 4
(\citealt{hodapp04a}), whose prototype telescope PS1 will probably see
first light in 2007, or the Large Scale Synoptic Telescope, which has
first light in 2014.  Searches for low surface brightness galaxies
(through their diffuse light) tend to be dominated by the scattered
light background, so it is difficult to anticipate how well of any of
these surveys can do. Any of these possibilities would require
spectroscopic follow-up --- probably searching for 21cm emission in
the radio or targeting HII regions in the galaxies.

Blind searches for galaxies in 21cm may show even more promise,
particularly since isolated dwarf galaxies virtually always exhibit HI
(\citealt{geha06b}) and since the dynamics of each galaxy will be
measured simultaneously with its detection.  Unfortunately, HIPASS
appears to not be deep enough to provide a competitive sample in this
respect (many of the SDSS galaxies in our sample are undetectable in
HIPASS; \citealt{geha06b}). ALFALFA (\citealt{giovanelli05a}) can
detect galaxies with HI masses of $10^7$ $h^{-2}$ $M_\odot$ at
distances of 20 $h^{-1}$ Mpc. If the full 7000 deg$^2$ planned survey
is completed, the overall volume mapped will be about $6\times 10^3$
$h^3$ Mpc$^{-3}$. That large a volume has an expected cosmic variance
of around 30\%, though given our restriction to isolated regions the
actual cosmic variance uncertainties will be lower.  Assuming a
baryonic to total mass fraction of $0.02$ (see Figure
\ref{isotf_mtobm}) and using the methods of \citet{bullock01a}, this
mass corresponds to $V_{\mathrm{max}} \sim 20$ km s$^{-1}$.  Galaxies
of this mass in the preliminary ALFALFA catalog of
\citet{giovanelli07a} have a median $W_{50}$ measurement of $\sim 40$
km s$^{-1}$, consistent with this estimate. Assuming (conservatively)
that 30\% cosmic variance errors dominate the uncertainties, using our
techniques here one could marginally exclude a warm dark matter
particle of 2 keV in mass. Future surveys such as MIRANDA will
contribute a similar volume (of a distinct chunk of the Universe), and
increase this precision somewhat. In any case, these new HI surveys
will push this technique into a low circular velocity regime that is
currently only tested with observations of Local Group satellites.

Even to make use of the current data, however, we probably require a
better understanding of the theoretical predictions. For example, as
we noted above, the excursion set predictions we are making for ``warm
dark matter'' are not entirely self-consistent, since we expect in
this regime that the hierarchical picture will start to break
down. Under these conditions, \citet{bode01a} found that the number of
forming halos was much smaller than the excursion set
prediction. Thus, it may be that our observations would put stronger
constraints on a correctly calculated prediction.  However, doing so
is difficult, since the effective mass resolution for warm dark matter
simulations appears to scale much less favorably than for cold dark
matter (see \citealt{wang07a}, who indeed argue that even the number
of halos predicted by \citealt{bode01a} is an overestimate).

Finally, it is worth noting two possible {\it fundamental} limits to
the technique we describe here. First, we rely on the 21cm emission to
probe the dynamics in the flat part of the rotation curve. While this
appears to be the case for most of our galaxies here (based on their
double-peaked morphology), it will not necessarily be true of the
typical very low mass galaxy. If so, one would need to resort to
comparing to cold dark matter predictions in the very inner parts of
halos.  Second, at lower masses, it may happen that reionization
evaporates the lowest mass halos, preventing the formation of stars
and the existence of any neutral gas at all.  If so, our counting
technique will fail.  For the galaxies in the mass range we study
here, this appears to be a weak effect, but it may occur at smaller
masses.

\section{ Summary }
\label{summary}

\begin{figure}
\figurenum{\fignum}
\plotone{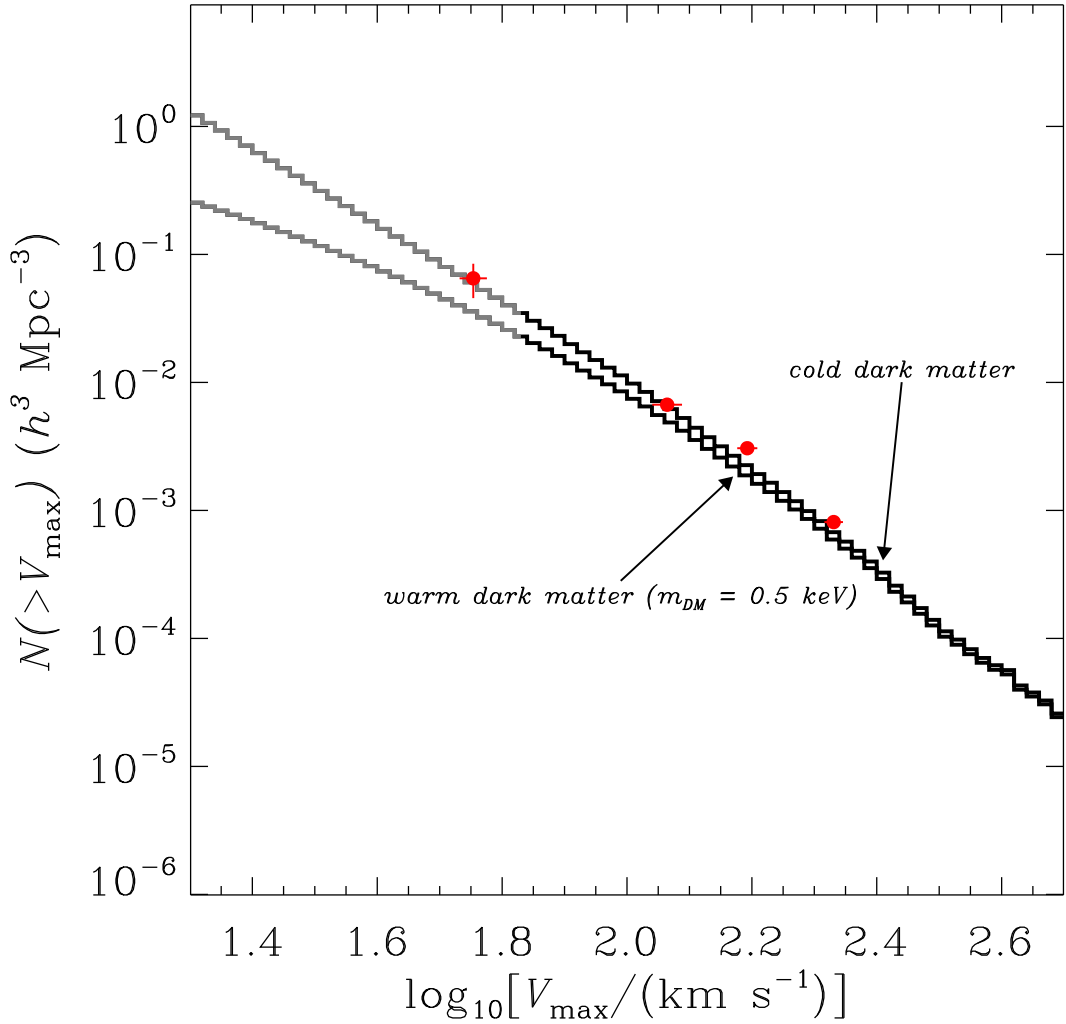}
\caption{\label{isovf_cvf_corr} Cumulative $V_{\mathrm{max}}$ function
	of isolated halos from Figure \ref{isovf_cvf}, corrected to the
	\citet{tegmark06a} cosmology (upper histogram) and that same
	cosmology with a light dark matter particle (lower histogram). }
\end{figure}

\begin{figure}
\figurenum{\fignum}
\plotone{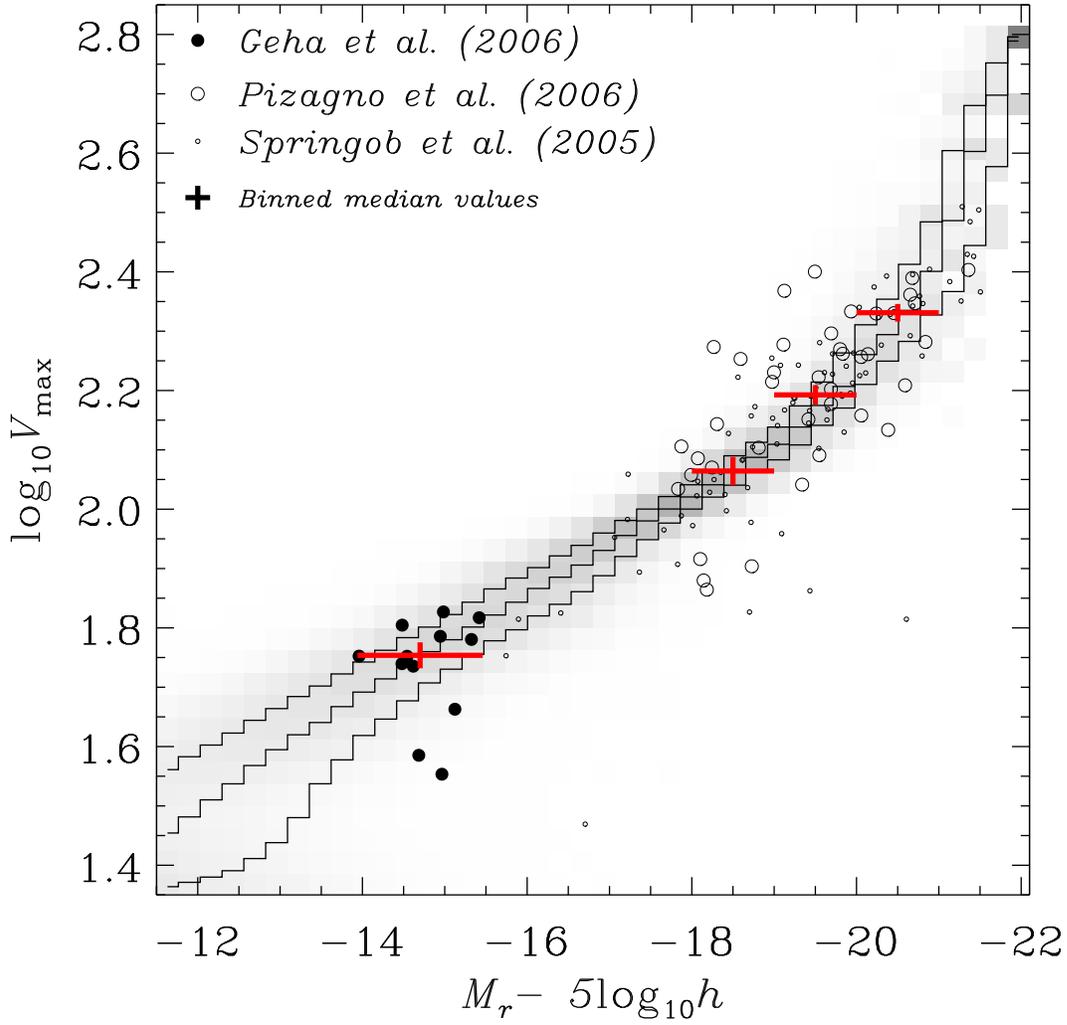}
\caption{\label{isotf} Conditional distribution of circular velocity
	as a function of absolute magnitude, to make our observed isolated
	galaxy luminosity function consistent with the predicted isolated
	halo circular velocity function. Lines are the quartiles of the
	distribution.  The overplotted points and binned median values are
	the data from Figure \ref{isotf_dataonly}.}
\end{figure}

\begin{figure}
\figurenum{\fignum}
\plotone{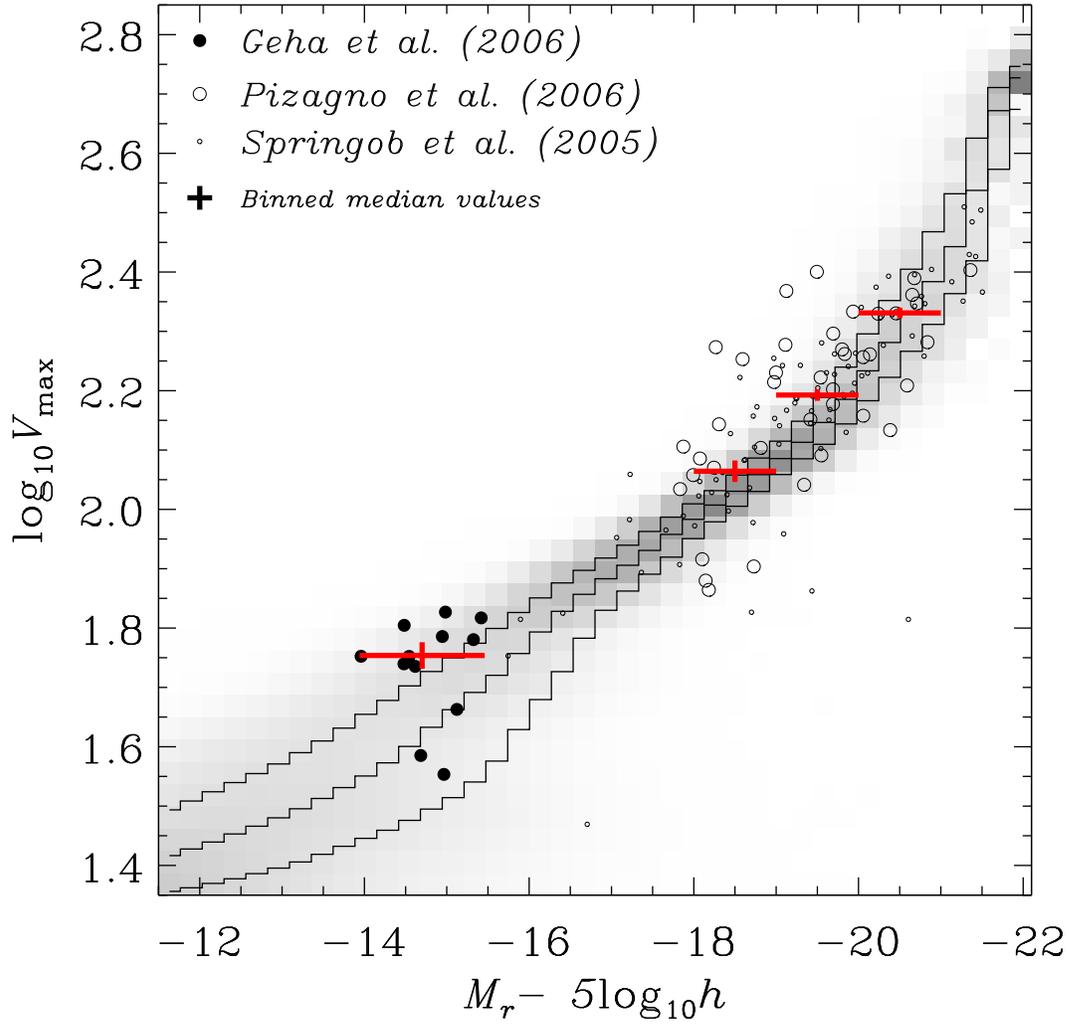}
\caption{\label{isotf_warm} Similar to Figure \ref{isotf} but based on
	the warm dark matter models.}
\end{figure}

\begin{figure}
\figurenum{\fignum}
\plotone{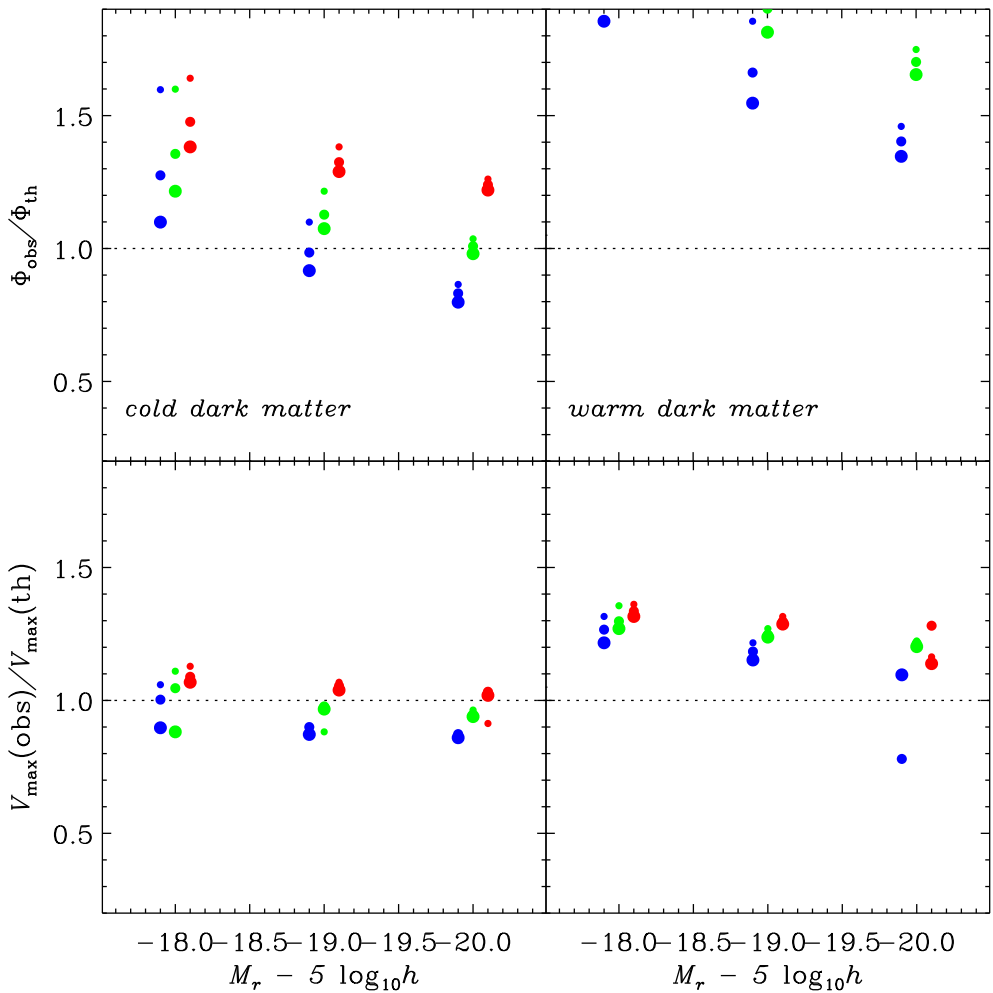}
\caption{\label{isotf_all} Dependence of our results on our definition
of ``isolated.'' The left panels refer to our cold dark matter model.
The right panels refer to the warm dark matter model with
$m_{DM}=\mdmlimit$ keV. The top panels show the ratio of the observed
number density of isolated galaxies with $M_r -5 \log_{10} h < -14.7$
to the predicted number of isolated halos with $V_{\mathrm{max}}> 56$
km s$^{-1}$ (that is, the vertical offset in Figure
\ref{isovf_cvf_corr}). The bottom panels show the ratio between the
median observed circular velocity at $M_r-5 \log_{10} h < -14.7$ to
that predicted in Figures \ref{isotf} and \ref{isotf_warm}. All of
these results are shown as a function of how we define ``isolated,''
which we do in 27 different ways here. We choose: (a) three different
absolute magnitude limits for the tracers in the observed sample, $M_r
-5\log_{10} h < -18$, $-19$, and $-20$, as shown by the rough
horizontal position; (b) for each sort of observed tracer, three
choices of minimum $V_{\mathrm{max}}$ for tracers in the predicted
sample, as listed in Table \ref{vmaxchoice} and as shown by the size
of the symbols (larger corresponds to higher circular velocity of
tracer); and (c) three different minimum projected radii from the
tracers, $r_P = 0.7$, $1.0$, and $1.3$ $h^{-1}$ Mpc, as shown by small
offsets in the horizontal positions (left to right).}
\end{figure}

We demonstrate that the predictions of cold dark matter are consistent
with the number density of isolated low circular velocity galaxies
($V_{\mathrm{max}} \sim 50$ km s$^{-1}$), at a precision of about 30\%
(Figures \ref{isovf_cvf_corr} and \ref{isotf}). Our major systematic
uncertainties (which dominate our error budget) are related to our
definition of ``isolated'' and our model of the surface brightness
completeness of the SDSS at low luminosities. These results represent
a valuable independent check of the small-scale predictions of the
cold dark matter model. Our technique avoids many of the systematic
uncertainties associated with observations of satellite galaxies in
the Local Group, related to the complex physics of ram pressure and
tidal stripping. From a statistical point of view, our results are
less powerful than the current constraints from the Ly$\alpha$ forest
power spectrum (\citealt{narayanan00a, abazajian06a, seljak06a}), and
only marginally exclude a dark matter particle with $m_{DM} \sim
\mdmlimit$ keV. With an improved sample, this technique will offer a
powerful and unique constraint at small scales.

\begin{figure}
\figurenum{\fignum}
\plotone{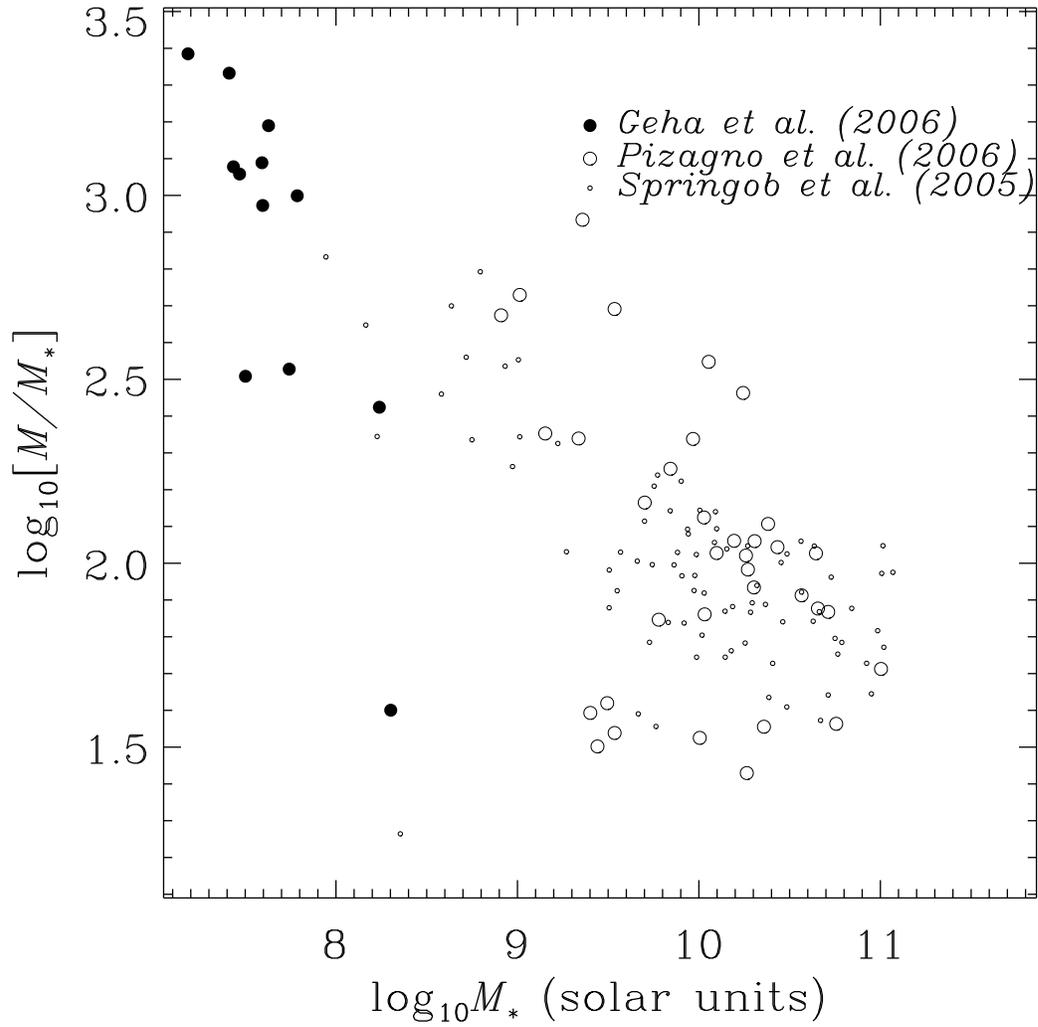}
\caption{\label{isotf_mtosm} Dynamical to stellar mass ratio as a
	function of stellar mass, for the three samples used in this paper
	(restricting to isolated galaxies with $b/a<0.5$).}
\end{figure}

We also find several secondary results that are relevant to the
formation of these dwarf galaxies:
\begin{enumerate}
\item At low luminosities, the Tully-Fisher relationship appears to be
	a function of environment, with dwarf satellite galaxies having
	lower circular velocities than isolated dwarf galaxies, by up to a
	factor of two or more. There is circumstantial evidence from the
	nature of the HI profiles that this effect is due to stripping of
	the outer gas in satellite galaxies, though other processes may be
	at work.
\item The baryonic mass fraction of galaxies (counting neutral gas
	plus stars) appears to be a weak function of luminosity down to $M_r
	- 5\log_{10} h \sim -14$, decreasing by 40\% at most (from about
	14\% to at minimum about 8\% the cosmic mean; Figure
	\ref{isotf_mtosm}). This result disfavors models which call for a
	preferentially large amount of baryonic outflow in dwarf galaxies
	(due to internal processes such as feedback).
\item The ratio of total to stellar mass is a very strong function of
	stellar mass, ranging from 50 or so at the highest luminosities to
	over 1000 at the lowest luminosities.
\end{enumerate}
Taken together with the deficit of neutral gas in dwarf galaxies near
luminous neighbors (\citealt{geha06b}), these results suggest that the
primary effects that remove gas and end star-formation in dwarf
galaxies (with circular velocites of $50$ km s$^{-1}$ or so) are
external interactions with bright neighbors, rather than internal
processes such feedback and outflows.

Increasing our sample of isolated dwarf galaxies with HI follow-up (it
is straightforward to increase the current sample by more than a
factor of five) would improve the precision of all of these results,
which are based on a relatively small sample of isolated galaxies.
Upcoming blind HI surveys such as the ALFALFA survey on Arecibo are
going to impose better constraints on cosmological models at small
scales, as well as better explore the issues of dwarf galaxy
formation. They will propel the study of field dwarf galaxies into a
low circular velocity and low mass regime previously studied only in
the Local Group.

\acknowledgments

We thank Andrey Kravtsov for sharing his simulation results, and
Andreas Berlind for his {\tt haloMF} software. For their interest,
encouragement, and good discussions, we thank Ari Maller, David
W.~Hogg, and Beth Willman. We thank Carl Bignell, Toney Minter, and
Karen O'Neil for help with the GBT observations. Partial support for
this work was provided by NASA-06-GALEX06-0300 and NSF-AST-0428465.
MG acknowledges support from a Plaskett Research Fellowship at the
Herzberg Institute of Astrophysics of the National Research Council of
Canada. AAW acknowledges the support of NSF grant 0540567.

Funding for the creation and distribution of the SDSS Archive has been
provided by the Alfred P. Sloan Foundation, the Participating
Institutions, the National Aeronautics and Space Administration, the
National Science Foundation, the U.S. Department of Energy, the
Japanese Monbukagakusho, and the Max Planck Society. The SDSS Web site
is http://www.sdss.org/.

The SDSS is managed by the Astrophysical Research Consortium (ARC) for
the Participating Institutions. The Participating Institutions are The
University of Chicago, Fermilab, the Institute for Advanced Study, the
Japan Participation Group, The Johns Hopkins University, the Korean
Scientist Group, Los Alamos National Laboratory, the
Max-Planck-Institute for Astronomy (MPIA), the Max-Planck-Institute
for Astrophysics (MPA), New Mexico State University, University of
Pittsburgh, University of Portsmouth, Princeton University, the United
States Naval Observatory, and the University of Washington.

\newpage

\clearpage
\clearpage

\setcounter{thefigs}{0}

\clearpage
\stepcounter{thefigs}

\clearpage
\stepcounter{thefigs}

\clearpage
\stepcounter{thefigs}

\clearpage
\stepcounter{thefigs}

\clearpage
\stepcounter{thefigs}

\clearpage
\stepcounter{thefigs}

\clearpage
\stepcounter{thefigs}

\clearpage
\stepcounter{thefigs}

\clearpage
\stepcounter{thefigs}

\clearpage
\stepcounter{thefigs}

\clearpage
\stepcounter{thefigs}

%\clearpage
%\stepcounter{thefigs}
%\begin{figure}
%\figurenum{\fignum}
%\plotone{isotf_mtol.ps}
%\caption{\label{isotf_mtol} blah}
%\end{figure}

\clearpage
\stepcounter{thefigs}

\newpage
\clearpage
\begin{deluxetable}{rrrr}
\tablewidth{0pt}
\tablecolumns{4}
\tablecaption{\label{vmaxchoice} Comparable circular velocities and
	absolute magnitudes}
\tablehead{ $M_r - 5\log_{10} h$ & $V_{\mathrm{max}}$ (km s$^{-1}$) &
  $V_{\mathrm{max}}$ (km s$^{-1}$) \cr
\cr
& (from TF) & (used) \cr}
\startdata
$-14.7$ & $56 \pm 3$ & --- \cr
$-18.0$ & $108 \pm 5$ & 95, 110, 125 \cr 
$-18.5$ & $116 \pm 6$ & --- \cr
$-19.0$ & $143 \pm 7$ & 125, 140, 155 \cr
$-19.5$ & $156 \pm 7$ & --- \cr
$-20.0$ & $180 \pm 4$ & 170, 180, 195 \cr
$-20.5$ & $214 \pm 7$ & --- \cr 
\enddata
%\startdata
%-18 & 117 & 97 & 87, 100, 112, 125 \cr
%-19 & 135 & 123 & 112, 125, 137, 150 \cr
%-20 & 182 & 162 & 150, 162, 175, 188 \cr
%\enddata
\tablecomments{ For each choice of $M_r$ (first column), this table
	yields the comparable $V_{\mathrm{max}}$ resulting from matching to
	the Tully-Fisher relation in Figure \ref{isotf_dataonly} (second
	column, ``TF''). It also lists the $V_{\mathrm{max}}$ values we
	actually tried in Figure \ref{isotf_all} (third column).}
\end{deluxetable}

\end{document}